\documentclass{jfm}

\usepackage{graphicx}
\usepackage{natbib}
        
\ifCUPmtlplainloaded \else
  \checkfont{eurm10}
  \iffontfound
    \IfFileExists{upmath.sty}
      {\typeout{^^JFound AMS Euler Roman fonts on the system,
                   using the 'upmath' package.^^J}%
       \usepackage{upmath}}
      {\typeout{^^JFound AMS Euler Roman fonts on the system, but you
                   dont seem to have the}%
       \typeout{'upmath' package installed. JFM.cls can take advantage
                 of these fonts,^^Jif you use 'upmath' package.^^J}%
      }
  \else
  \fi
\fi

\ifCUPmtlplainloaded \else
  \checkfont{msam10}
  \iffontfound
    \IfFileExists{amssymb.sty}
      {\typeout{^^JFound AMS Symbol fonts on the system, using the
                'amssymb' package.^^J}%
       \usepackage{amssymb}%

      }{}
  \fi
\fi

\ifCUPmtlplainloaded \else
  \IfFileExists{amsbsy.sty}
    {\typeout{^^JFound the 'amsbsy' package on the system, using it.^^J}%
     \usepackage{amsbsy}}
    {}
\fi

\newsavebox{\astrutbox}
\sbox{\astrutbox}{\rule[-5pt]{0pt}{20pt}}

\usepackage{amsmath}
\usepackage{multirow}

\title{Alignment of vorticity and rods with Lagrangian fluid stretching in turbulence}

\author[R. Ni and N. T. Ouellette and G. A. Voth]%
{RUI NI$^{1,2}$, \ns NICHOLAS T. OUELLETTE$^{2}$ \break and GREG A. VOTH$^{1}$%
  \thanks{Email address for correspondence: gvoth@wesleyan.edu}}

\affiliation{$^1$Department of Physics, Wesleyan University, Middletown, Connecticut 06459, USA
\\[\affilskip]
$^2$Department of Mechanical Engineering $\&$ Materials Science, Yale University, New Haven, Connecticut 06520, USA}

\begin{document}
\maketitle



\begin{abstract}
Stretching in continuum mechanics is naturally described using the Cauchy-Green strain tensors. These tensors quantify the Lagrangian stretching experienced by a material element, and provide a powerful way to study processes in turbulent fluid flows that involve stretching such as vortex stretching and alignment of anisotropic particles. Analyzing data from a simulation of isotropic turbulence, we observe preferential alignment between anisotropic particles and vorticity. We show that this alignment arises because both of these quantities independently tend to align with the strongest Lagrangian stretching direction, as defined by the maximum eigenvector of the left Cauchy-Green strain tensor. In particular, anisotropic particles approach almost perfect alignment with the strongest stretching direction. The alignment of vorticity with stretching is weaker, but still much stronger than previously observed alignment of vorticity with the eigenvectors of the Eulerian strain rate tensor. The alignment of strong vorticity is almost the same as that of rods that have experienced the same stretching. 
\end{abstract}

\section{Introduction}
Stretching of a fluid element in a turbulent flow is a dynamic process that is naturally expressed in the Lagrangian framework. For three-dimensional flows, this process can be visualized by considering a sphere that is distorted into a tri-axial ellipsoid as it is stretched by the flow.  In continuum mechanics, the deformation of an object  is commonly described by the deformation gradient tensor $F_{ij}=(\partial x_i/\partial X_j)$, where $\bf{X}$ is an initial position and ${\bf{x}}$ is a final position.  After a sphere is deformed into an ellipsoid, the orientations of the three principle axes of the ellipsoid are given by the eigenvectors of the left Cauchy-Green strain tensor ${\bf{C}}^{(L)}={\bf{FF}}^T$\citep{Malvern}. 

In studies of turbulence, the most widely used properties of the Cauchy-Green strain tensor have been its eigenvalues, which specify the lengths of the principle axes and therefore the shape of the ellipsoid~\citep{1990JFMGirimaji, 2005JFMLuthi, 2006PREGuala}.  The eigenvalues are directly related to the Lyapunov exponents (also known as the  stretching rates) \citep{2006POFBec, 1993JASPierrehumbert}.  One application that uses the eigenvalues has been identification of Lagrangian coherent structures \citep{2007JFMGreen,2013PTPeacock}, which provide insights into mixing and transport.     

In this paper, we show that the eigenvectors of the Cauchy-Green strain tensor provide a Lagrangian basis in which the alignment of both passive vectors and vorticity is remarkably simple.   Passive vectors along with thin rods and material line segments become preferentially aligned with the longest principle axis of the ellipsoid, and at long times approach perfect alignment with the eigenvector corresponding to maximum stretching.    Vorticity is more complex since it is an active vector that is both amplified by stretching and affected by viscosity.   We find that the vortex stretching process also leads to strong alignment with the direction of maximum stretching, but that the degree of alignment saturates after 10 $\tau_{\eta}$, where $\tau_{\eta}$ is the Kolmogorov time scale.

  There is an extensive literature on the alignment of both passive vectors and vorticity in turbulence. In many applications, orientation dynamics reduce to the passive vector problem. These include the orientation of thin rods~\citep{2011POFShima, 2011NJPPumir,2013POFMehlig,2013PRLMehlig}, material line segments \citep{1991JFMDresselhaus, 2005JFMLuthi}, and magnetic field lines in a medium with high conductivity~\citep{monin}.   \citet{1952PRSLABatchelor} provided theoretical predictions for material line and surface stretching based on the assumption of persistent straining over a short period of time. Subsequently, both simulations \citep{1990JFMGirimaji} and experiments \citep{2005JFMLuthi, 2006PREGuala} have studied the stretching of material lines by focusing on the alignment of a material line with the eigenvectors of the Eulerian strain rate tensor ($\hat{\bf{e}}_i$, $i=1,2,3$). Both of these studies used the eigenvalues of the Cauchy-Green tensor to determine the deformation of a material volume;  however, neither made connections between the orientation of material line and the eigenvectors of the Cauchy-Green strain tensor. \citet{2011NJPPumir} and \citet{2012JPAMTWilkinson} focused on the relative orientation of rods with respect to the vorticity vector as well as the Eulerian strain rate tensor.  They found that the rods align more strongly with the vorticity and identified that the stretching term in the equations of motion for both vorticity and rods is responsible for the alignment.  As we explain below,  the effect of the stretching term is to align both  passive vectors and the vorticity vector with the largest Lagrangian stretching direction.


In turbulence, \citep{1938PRSLATaylor} conjectured that the turbulent cascade mechanism relies on the amplification of vortices by stretching, which subsequently leads to breakup of large vortices into smaller ones. Most previous studies of vortex stretching have considered the alignment of vorticity ${\bf{\omega}}$ with the eigenvectors of the Eulerian strain rate tensor  \citep{1987POFAshurst, 1996POFHuang}, which give the instantaneous stretching directions. It was found that the instantaneous vorticity tends to align with the intermediate eigenvector $\hat{\bf{e}}_2$ \citep{1987POFAshurst}, and that the tendency for this alignment increases with increasing magnitude of the vorticity $|\omega|$ \citep{1996POFHuang}.  This finding is often argued to be counterintuitive because the vorticity should be preferentially aligned with largest eigenvector due to conservation of angular momentum \citep{2011NaturePXu}. One possible explanation for these observations is that the alignment is caused by self-induced stretching \citep{1990NatureShe}, implying that the strain field in strongly vortical regions is dominated by the vorticity itself. A similar kinematic argument is that a strong vortex tube behaves like a two-dimensional flow, with the most extensive and compressive strains lying in the equatorial plane and leaving the intermediate eigenvector to be aligned with the vortex \citep{1992POFJimenez}.  Alternatively, \citet{2008POFHamlington}  approached the alignment of vorticity with the strain rate eigenvectors by decomposing the strain rate into local and nonlocal components.   They showed that the vorticity tends to align with the most extensional eigenvector of the nonlocal component of the strain rate tensor even though it aligns with the intermediate eigenvector of full strain rate tensor. 

In addition to studies that focus on the instantaneous alignment, \citet{2011NaturePXu} and \citet{2013POFAlain} have more recently provided new insight into this problem by considering the alignment between the eigenvectors of the strain rate tensor at a given time $t_0$ with the vorticity at a \emph{later} time $t=t_0+\Delta t$. They found that the alignment between ${\bf{\omega}}(t)$ and $\hat{\bf{{e}}}_1(t_0)$ grows with $\Delta t$ for $\Delta t<2\tau_{\eta}$. By including time in the analysis, these results begin to include dynamical information about the stretching process. To build on this insight, we consider the stretching that the vorticity has experienced in a fully Lagrangian way rather than only at two time instants.

The full Lagrangian dynamics of the Eulerian velocity gradient tensor have been extensively modeled and studied \citep{2011ARFMMeneveau}. Most of those studies have focused primarily on how to model its Lagrangian evolution. Note, however, that the Cauchy-Green strain tensors that encode the Lagrangian stretching are obtained by integrating the Lagrangian velocity gradient over a finite time interval. Thus, Lagrangian stretching and the Lagrangian velocity gradient are distinct (though related) concepts. \citet{2007POFLi} analytically obtained the Lagrangian stretching from the Restricted Euler model, and found good agreement with results from a direct numerical simulation of the Navier--Stokes equations. But the alignment statistics of the Lagrangian stretching with the vorticity or any other vectors have not been considered.

In this paper, we demonstrate that once the stretching has been appropriately defined in a fully Lagrangian way, the geometric properties of vortex stretching and the orientation dynamics of anisotropic particles become simple and intuitive. The paper is organized as follows: In \S \ref{sec:method}, we give a brief discussion of the definition of Lagrangian stretching and numerical method we use to determine Lagrangian stretching. In \S \ref{sec:results}, we present results on the alignment of rods and vorticity with the Lagrangian stretching, and compare them with other definitions of stretching. In \S \ref{sec:rods}, we discuss the dynamics of rod orientation in detail. We also derive analytical bounds for the alignment, and compare our predictions with our numerical results. The Lagrangian stretching of vortices with different magnitudes is explored in detail in \S \ref{sec:vortex}, where we also discuss the correlation between the magnitudes of vorticity and stretching.

\section{Methods}
\label{sec:method}
In this paper, we study the Lagrangian stretching as defined by the Cauchy-Green strain tensor by analyzing data from a direct numerical simulation (DNS) of homogeneous isotropic turbulence~\citep{2009PREBenzi}. The data were generated from a simulation with $N^3=512^3$ collocation points, corresponding to a Taylor-microscale Reynolds number of $R_{\lambda}=180$. A total of $7\times 10^4$ Lagrangian trajectories were followed for $O(1)$ large-eddy turnover times, and the velocity gradient tensor at the tracer positions was stored. The orientations of rod-shaped Lagrangian tracers with an aspect ratio of $20$ were obtained by integrating Jeffery's equation \citep{1922PRSLAJeffery} along each trajectory \citep{2012PRLShima}. This choice of aspect ratio is arbitrary; we find, however, that the orientation dynamics for rods with aspect ratio larger than about 10 are insensitive to the aspect ratio, as they behave essentially as material-line segments.

For completeness, we briefly discuss here the Cauchy-Green strain tensors, their eigenvectors, and how we compute them.  More details can be found in continuum-mechanics textbooks \citep{Chadwick, Malvern}. Consider an infinitesimal spherical fluid element at some time $t_0$. After a time $\Delta t$, it will in general have been stretched into an ellipsoid. The position of any point $\bf{X}$ inside the spherical element at $t_0$ will be mapped to a position ${\bf{x}}$ inside the ellipsoid at $t=t_0+\Delta t$. We can define a deformation gradient tensor that characterizes the deformation experienced by the fluid element as $F_{ij}=(\partial x_i/\partial X_j)$. 
${\bf F}$ evolves as $dF_{ij}(t)/dt=A_{ik}(t)F_{kj}(t)$, where ${\bf A}$ is the instantaneous velocity gradient, with the initial condition $F_{ij}(t_0)=\delta_{ij}$. We obtain ${\bf{F}}(t)$ from the DNS data by integrating over $\Delta t$ using a fourth-order Runge-Kutta scheme.  

Any (affine) deformation can be expressed as pure stretching followed by a rotation or as a rotation followed by pure stretching. Thus, the deformation gradient tensor can be decomposed as $\bf{F} = \bf{RU} = \bf{VR}$, where $\bf{R}$ is an orthogonal rotation tensor and $\bf{U}$ and $\bf{V}$ are, respectively, the right and left stretch tensors.  The stretching, without any contribution from rotation, can be obtained from the two symmetric inner products of $\bf{F}$ with itself:
\begin{eqnarray}
{\bf{C}}^{(L)}&={\bf{F}}{\bf{F}}^T=\bf{VR R^T V^T}={\bf{V}}^2 \\
{\bf{C}}^{(R)}&={\bf{F}}^T{\bf{F}}=\bf{U^T R^T  R U}={\bf{U}}^2 
\end{eqnarray}
${\bf{C}}^{(L)}$ and ${\bf{C}}^{(R)}$ are, respectively, the left and right Cauchy-Green strain tensors.   
These two tensors have the same eigenvalues $\Lambda_{i}$  (i=1,2,3), but different eigenvectors $\hat{\bf{e}}_{Li}$ (left) and $\hat{\bf{e}}_{Ri}$ (right).   The largest eigenvalue,  $\Lambda_1 > 1$, indicates extension, the smallest eigenvalue, $\Lambda_3 < 1$, indicates contraction, and the intermediate eigenvalue can indicate either extension or contraction. 

The physical meaning of the eigenvectors of the two Cauchy-Green strain tensors can be shown in a few simple steps. Consider a material line segment that is {\it{initially}} aligned with largest right eigenvector, ${\bf{l}}(t_0)=\hat{\bf{e}}_{R1}$. After some $\Delta t$, the material line will be deformed into ${\bf{l}}(t)={\bf{F}} {\hat{\bf{e}}}_{R1}$. Multiplying both sides of the equation with $C^{(L)}$, we have 
\begin{equation}
{\bf{C}}^{(L)} {\bf{l}}(t)= {\bf{C}}^{(L)} [{\bf{ F \hat{\bf{e}}}}_{R1}]= {\bf{F F^T F}} \hat{\bf{e}}_{R1}= {\bf{F C}^{(R)}} {\hat{\bf{e}}}_{R1}= {\bf{F}} \Lambda_1 \hat{\bf{e}}_{R1}=\Lambda_1 {\bf l}(t).
\end{equation}
Thus, the \textit{final} direction of the material line, $\hat{\bf{l}}(t)={\bf{l}}(t)/|{\bf{l}}(t)|$, is the eigenvector of the left Cauchy-Green strain tensor that corresponds to the maximum eigenvalue---namely $\hat{\bf{e}}_{L1}$. The same proof applies for the other two pairs of eigenvectors ($\hat{\bf{e}}_{R2}$, $\hat{\bf{e}}_{L2}$ and $\hat{\bf{e}}_{R3}$, $\hat{\bf{e}}_{L3}$).  Material lines that initially align with an eigenvector of the right tensor end up aligned with the counterpart eigenvector of the left tensor. Hereafter, to capture their physical meaning, the $\hat{\bf{e}}_{R1}$ and $\hat{\bf{e}}_{L1}$ are also referred to as the initial and final largest Lagrangian stretching directions, respectively. Similarly, $\hat{\bf{e}}_{1}$ at time $t_0$ and $t$ are referred to as the initial and final largest Eulerian stretching directions.

\section{Results}
\label{sec:results}

\subsection{Average alignments of rods and vorticity}

\begin{figure}
\begin{center}
$\begin{array}{cc}
\includegraphics[width=5in]{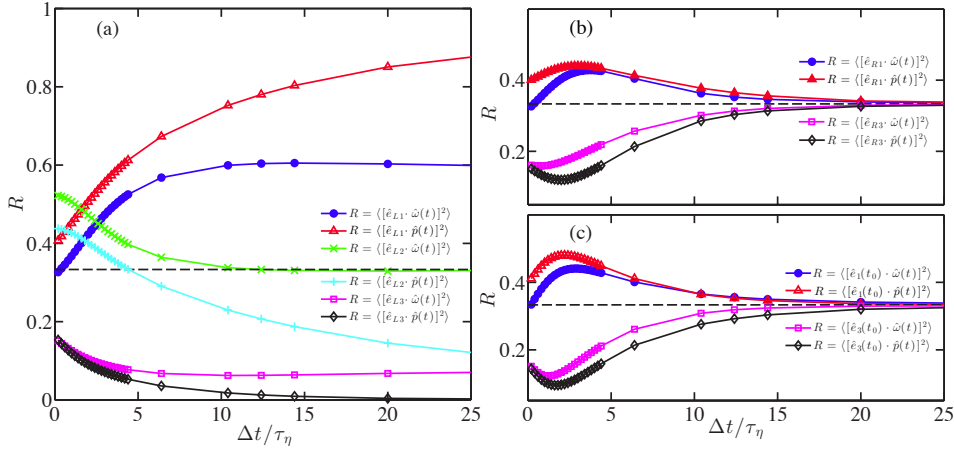}
\end{array}$
\caption{(Color online) The alignment of rods ($\hat{\bf{p}}(t)$) and vorticity ($\hat{\bf{\omega}}(t)$) with respect to different definitions of the stretching: (a) the eigenvectors of the left Cauchy-Green strain tensor $\hat{\bf{e}}_{Li}$; (b) the eigenvectors of the right Cauchy-Green strain tensor $\hat{\bf{e}}_{Ri}$; and (c) the eigenvectors of the strain rate tensor at the initial time $\hat{\bf{e}}_{i}(t_0)$. For all three panels, $i=1,2,3$ are the indices for the eigenvectors that corresponding to the largest, intermediate, and smallest eigenvalues of the tensor of interest, and the horizontal dashed lines show $R=1/3$,  corresponding to the alignment between two randomly oriented vectors.}
\label{fig:omgcg}
\end{center}
\end{figure} 

Figure \ref{fig:omgcg} shows the alignment of infinitesimal rods $\hat{\bf{p}}(t)$ and the vorticity $\hat{\bf{\omega}}(t)$ with both the Lagrangian and Eulerian stretching directions. The alignment is quantified using the square of the cosine of the angles between two unit vectors. Both rods and vorticity align most strongly with the final largest Lagrangian stretching direction, $\hat{\bf{e}}_{L1}$, especially if we use time intervals of at least $\Delta t=10 \tau_{\eta}$ to calculate $\hat{\bf{e}}_{L1}$. 

In Fig.~\ref{fig:omgcg}(a), the degree of alignment of rods with $\hat{\bf{e}}_{L1}$ is higher than it is for the vorticity because infinitesimal rods are material line segments that passively align with $\hat{\bf{e}}_{L1}$. The evolution of vorticity, on the other hand, is more complicated. From its equation of motion, the vorticity evolves both due to stretching by the velocity gradient tensor and to the tearing or reconnection at small scales that can be induced by viscosity. The stretching of vorticity is the same as the stretching of a material line segment, and will tend to align the vorticity in the same direction as $\hat{\bf{e}}_{L1}$. Indeed, Fig.~\ref{fig:omgcg}(a) shows that the alignment between $\hat{\bf{\omega}}(t)$ and $\hat{\bf{e}}_{L1}$ is very strong, but that this alignment reaches a plateau at 0.6 after 10 $\tau_{\eta}$. We interpret this plateau as a result of the dynamic balance between stretching, which moves $\hat{\bf{\omega}}(t)$ toward $\hat{\bf{e}}_{L1}$, and viscous effects, which move them apart. For the same reason, $\hat{\bf{\omega}}$ cannot be perfectly perpendicular to $\hat{\bf{e}}_{L3}$ and $\hat{\bf{e}}_{L2}$, as also seen in Fig.~\ref{fig:omgcg}(a). We remark that, since both rods and vorticity independently show strong alignment with $\hat{\bf{e}}_{L1}$ due to the stretching process, they must also align with each other~\citep{2011NJPPumir}.


At $\Delta t=0$, Fig.~\ref{fig:omgcg} tells us that the instantaneous alignment between rods ($\hat{\bf{p}}$) and the eigenvectors of the Eulerian strain-rate tensor $\hat{\bf{e}}_{i}$ are 0.40, 0.44, and 0.16 for $i = (1,2,3)$. These values indicate that rods are slightly more aligned with $\hat{\bf{e}}_{2}$ than $\hat{\bf{e}}_{1}$, consistent with some previous work \citep{2011NJPPumir}. However, other studies have reported different results. Experimentally, \citet{2005JFMLuthi} and \citet{2006PREGuala} found that material lines with random initial orientations align preferentially with $\hat{\bf{e}}_1$ after $\sim6\tau_{\eta}$. \citet{2008thesisWan} studied the alignment of material lines at six different Reynolds numbers, ranging from $R_\lambda = 17$ to 430. He found that at short times ($t<10 \tau_{\eta}$), material lines were preferentially oriented along $\hat{\bf{e}}_1$, a finding consistent with the experimental results \citep{2005JFMLuthi, 2006PREGuala}. But in the long-time limit ($t>10\tau_{\eta}$), the alignment of material lines was very sensitive to the Reynolds number. At $R_{\lambda}=17$ and 430, material lines aligned more with $\hat{\bf{e}}_1$, while at intermediate Reynolds numbers ($R_{\lambda}=50$, 73, 120 and 240), material lines aligned better with $\hat{\bf{e}}_2$.  In all of these cases, however, the observed alignment of material lines with any of the $\hat{\bf{e}}_i$ was much weaker than the alignment we observe with the Lagrangian stretching direction.

As time evolves, the alignment between the vorticity, $\hat{\bf{\omega}}$, and the eigenvectors of the left Cauchy-Green tensor $\hat{\bf{e}}_{Li}$ changes from 0.32, 0.52, and 0.16 at $\Delta t = 0$ for $i = (1,2,3)$, eventually saturating at 0.61, 0.33, and 0.06 at $\Delta t=15\tau_{\eta}$. This evolution indicates that if we define stretching in a Lagrangian way rather than an Eulerian one, the vorticity aligns with the largest stretching direction rather than the intermediate eigenvector. Numerous studies have proposed explanations for the puzzling alignment between vorticity and the intermediate eigenvector of the Eulerian strain rate \citep{1987POFAshurst, 1996POFHuang, 2008POFHamlington,2011NaturePXu}; in a fully Lagrangian description, however, the alignment is much simpler. The vorticity becomes preferentially aligned with the largest stretching direction due to angular momentum conservation, just as one would intuitively expect. 

The alignment trends between both rods and the vorticity and $\hat{\bf{e}}_{R1}$ and $\hat{\bf{e}}_{1}(t_0)$ are very similar to each other, as shown in Fig.~\ref{fig:omgcg}(b) and (c). To explain this similarity, we recall first that $\hat{\bf{e}}_{R1}$ gives the direction in which an initial spherical fluid element will be most strongly stretched after $\Delta t$. Similarly, $\hat{\bf{e}}_{1}(t_0)$ is the direction of strongest stretching at the initial time $t_0$ \citep{2013POFAlain}. Thus, the alignment of $\hat{\bf{\omega}}(t)$ and $\hat{\bf{p}}(t)$ with either $\hat{\bf{e}}_{R1}$ or $\hat{\bf{e}}_1(t_0)$ arises from the same dynamical picture. In each case, alignment with the initially strongest stretching direction builds up over a short but finite time. But as $\Delta t$ grows, the direction of strongest stretching at the initial time becomes more and more uncorrelated with the final orientation of the rods or vorticity. Thus, all the curves in Fig.~\ref{fig:omgcg}(b) and (c) approach $1/3$ (the value expected for randomly oriented vectors) in the long time limit.

\subsection{Rods: slow approach to perfect alignment}
\label{sec:rods}

Figure \ref{fig:omgcg}(a) shows that within $15\tau_{\eta}$, rods become almost perfectly perpendicular to $\hat{\bf{e}}_{L3}$ and so lie in the plane $S_{12}$ containing $\hat{\bf{e}}_{L1}$ and $\hat{\bf{e}}_{L2}$. Subsequently, over a much longer time scale, they become parallel to $\hat{\bf{e}}_{L1}$ and perpendicular to $\hat{\bf{e}}_{L2}$.  
To understand this slow alignment, we can again visualize stretching as the process of deforming a sphere into an ellipsoid with principle axes of length $l_i=\sqrt{\Lambda_i}$ (i=1,2,3). Note that since the flow is incompressible, $l_1l_2l_3=1$ and the stretching can be specified using only two independent parameters.  

\begin{figure}
\begin{center}
$\begin{array}{cc}
\includegraphics[width=2.6in]{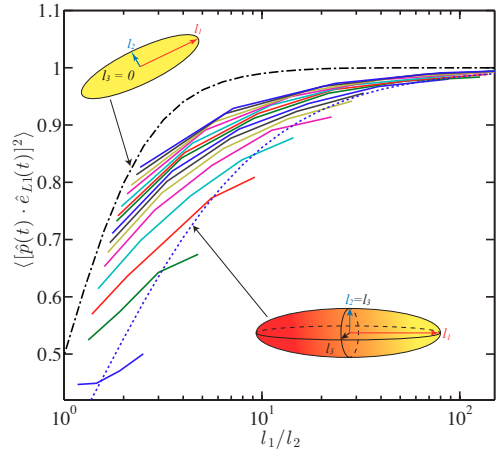}
\end{array}$
\caption{(Color online) The alignment of the orientation of rods $\hat{\bf{p}}(t)$ with respect to the largest Lagrangian stretching direction $\hat{\bf{e}}_{L1}(t)$ as a function of $l_1/l_2$. From bottom to top, the solid lines represent different $\Delta t$, spaced linearly from 1 $\tau_{\eta}$ to 15 $\tau_{\eta}$. The lower dotted line and the upper dash-dotted line show the ideal cases where a spherical fluid element has been stretched into an axisymmetric ellipsoid ($l_2=l_3$) and a flat two-dimensional ellipsoid ($l_3=0$) respectively.}
\label{fig:ratio}
\end{center}
\end{figure}

In Fig.~\ref{fig:ratio}, the alignment between $\hat{\bf{p}}(t)$ and $\hat{\bf{e}}_{L1}(t)$ is plotted as a function of $l_1/l_2$ for different $\Delta t$, ranging from 1 $\tau_{\eta}$ to 15 $\tau_{\eta}$. For all $\Delta t$, the alignment increases monotonically with $l_1/l_2$, suggesting that the geometrical aspect ratio $l_1/l_2$ controls the orientation for the rods in the plane $S_{12}$. The extension of each solid line give us an idea of the width of the $l_1/l_2$ distribution. Both the mean value and the range of the ratio $l_1/l_2$ increase with increasing $\Delta t$. At $\Delta t$ = 15$\tau_{\eta}$, $l_1/l_2$ varies over two orders of magnitude, from 2 to more than 200. The alignment for large $l_1/l_2\sim100$ is almost perfect; but there are still a non-negligible number of samples with small $l_1/l_2<10$ so that the overall alignment with $\hat{\bf{e}}_{L1}(t)$ is imperfect. 
When visualizing stretching in 3D turbulence, it is useful to consider two limiting cases: deformation into pancake shapes ($l_1\approx l_2\gg l_3$) and into cigar shapes ($l_1\gg l_2 \approx l_3$) \citep{1990JFMGirimaji}. For pancakes, the extreme case is a two-dimensional ellipsoid with $l_3$ approaching zero. In this case, rods will have lost all orientational freedom in the $\hat{\bf{e}}_{L3}$ direction, which makes it more likely that they will be aligned with $\hat{\bf{e}}_{L1}(t)$. Thus, the thin pancake limit should be the upper bound for all curves for a given value of $l_1/l_2$. To calculate the expected alignment in this case, we used a model proposed for two-dimensional flow \citep{2011POFShima}; the results are shown with the dash-dotted line in Fig. \ref{fig:ratio}. 
For cigars, which are axisymmetric ellipsoids with $l_2=l_3$, rods have the most freedom to align in the $\hat{\bf{e}}_{L3}$ direction of all shapes with a given value of $l_1/l_2$. Their alignment with $\hat{\bf{e}}_{L1}$, therefore, will be the smallest as compared with other shapes. Here, we provide a simple analytic model to calculate this effect.  Consider a radial material line in a unit sphere whose orientation is $(x,y,z)$ in the coordinate system $\{\hat{\bf{e}}_{Li}\}$.   After $\Delta t$, the unit sphere will be stretched into an ellipsoid with principle axes $l_{i}$. The material line will be mapped to the corresponding orientation in the ellipsoid, pointing along $(l_1x,l_2y,l_3z)$. Given the two extra conditions $l_1l_2l_3=1$ (incompressible flow) and $l_2=l_3$ (an axisymmetric ellipsoid), we have
\begin{equation}
\langle[\hat{\bf{p}}(t)\cdot\hat{\bf{e}}_{L1}]^2\rangle=\int_{-1}^{1}\frac{l_1^2x^2}{l_1^2x^2+l_2^2y^2+l_3^2z^2}P(x)dx=\frac{1}{2}\int_{-1}^{1}\frac{s^2}{s^2+(\frac{1}{x^2}-1)}dx, 
\label{eq:int}
\end{equation}
where $s=l_1/l_2$ is the aspect ratio of the ellipsoid. $P(x)$ is the probability density function (PDF) of $x$. If we assume that the initial orientation of the material line inside the sphere is uniformly distributed, then $P(x)=1/2$ for $x\in[-1,1]$, and we obtain the dotted line in Fig. \ref{fig:ratio}. If the rods were initially randomly oriented, this would be the lower bound for their alignment.  In the simulation, however, we evolve rods with the turbulence until they reach a steady state before measuring their orientational statistics, in order to obtain results that are closer to the experimentally measurable case of advected rods~\citep{2012PRLShima}.   This steady state has a non-random initial orientation, which leads to alignments that are sometimes lower than the axisymmetric limit shown. Note that there are fewer points below the dotted line as $\Delta t$ increases, since the initial conditions become less and less relevant for larger $\Delta t$.

\subsection{Vortex stretching: effects of vorticity and stretching magnitudes}
\label{sec:vortex}
\begin{figure}
\begin{center}
$\begin{array}{cc}
\includegraphics[width=2.6in]{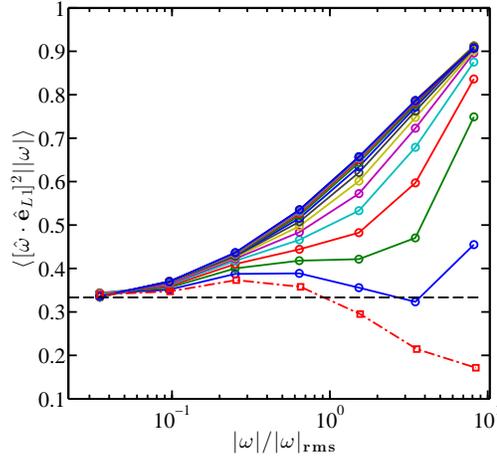}
\end{array}$
\caption{(Color online) The alignment of $\hat{\bf{\omega}}(t)$ with $\hat{\bf{e}}_{L1}$ conditioned on the magnitude of vorticity $|\omega(t)|$ for different $\Delta t$, spaced linearly from 0 $\tau_{\eta}$ (bottom dash-dotted line) to 15 $\tau_{\eta}$ (top solid line) with time step 1 $\tau_{\eta}$. Note that the $\hat{\bf{e}}_{L1}$ at $\Delta t=0$ is equal to $\hat{\bf{e}}_1(t)$.
 }
\label{fig:condw}
\end{center}
\end{figure}
 Part of the reason for the imperfect alignment of vorticity with Lagrangian stretching in Fig. \ref{fig:omgcg} is that the result is averaged over all vorticity magnitudes.    In Fig. \ref{fig:condw}, we show the alignment between $\hat{\bf{\omega}}(t)$ and $\hat{\bf{e}}_{L1}$ conditioned on the vorticity magnitude for different $\Delta t$. Since $\hat{\bf{e}}_{L1} = \hat{\bf{e}}_{1}(t)$ for $\Delta t=0$, the bottom dash-dotted line also shows the conditional alignment between $\hat{\bf{\omega}}(t)$ with $\hat{\bf{e}}_{1}(t)$, the largest eigenvector of the Eulerian strain rate. In the Eulerian case, the results are very complicated: large vorticity is preferentially oriented perpendicular to the largest stretching direction, while small vorticity shows weak alignment. But, as with the cases discussed above, the physical picture becomes much clearer if we work with Lagrangian stretching. In the Lagrangian case, for $\Delta t> 10 \tau_{\eta}$, all curves collapse with each other; thus, there is a well-defined asymptotic alignment.  
The asymptotic curve is reached in the same time range where the overall alignment between $\hat{\bf{\omega}}(t)$ and $\hat{\bf{e}}_{L1}$ in Fig \ref{fig:omgcg}(a)  reaches its plateau.   The positive slope of the asymptotic curve tells us simply that stronger vorticity is better aligned with the largest Lagrangian stretching direction. In particular, we note that the alignment reaches 0.9 for $|\omega|=8.5|\omega|_{rms}$ for which the vortex structures are known to be predominantly tubular \citep{1990NatureShe}. This alignment is very high, particularly when compared with the alignment averaged over all vorticity magnitudes. We observe this same alignment of 0.9 for rods if we condition on the same vorticity magnitude; thus, large vorticity behaves just like a rod, and any effect of viscosity on its alignment is negligible. For smaller vorticity, the alignment is smaller due to viscous effects.  For the weakest vorticity ($|\omega|<|\omega|_{rms}/10$), the relative orientations of $\hat{\bf{\omega}}$ and $\hat{\bf{e}}_{L1}$ are purely random.

\begin{figure}
\begin{center}
$\begin{array}{cc}
\includegraphics[width=4in]{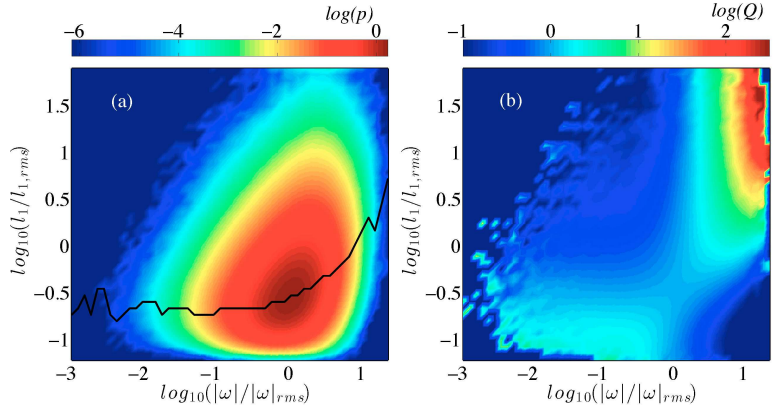}
\end{array}$
\caption{(Color online) (a) Joint PDF of the vorticity magnitude $|\omega|$ and amount of stretching $l_1$ normalized by their  own standard deviations at $\Delta t=10 \tau_{\eta}$. The black solid line shows the the most probable stretching $l_1$ for each $|\omega|$. (b) the PDF quotients Q (Eq. \ref{eq:q}) for the same quantities.}
\label{fig:jpdf}
\end{center}
\end{figure}

In the vortex-stretching process,  the magnitude of the stretching must also play an essential role. Figure \ref{fig:jpdf}(a) shows the joint PDF between $|\omega|$ and $l_1$, both of which are normalized by their own standard deviations. The black solid line shows the most probable value of $l_1$ at each $|\omega|$. For vorticity magnitudes smaller than $|\omega|_{rms}$, the most probable stretching is relatively small and does not change much with $|\omega|$; but for $|\omega|>|\omega|_{rms}$, it increases very quickly with $|\omega|$. Large vorticity is likely to occur simultaneously with large stretching. To see the correlation between these two variables in another way, we also plot in Fig.~\ref{fig:jpdf}(b) the PDF quotients $Q$, defined as \citep{2007PRLXu}
\begin{equation}
Q(|\omega|,\Lambda_{L1})\equiv\frac{P(|\omega|,\Lambda_{L1})}{P(|\omega|)P(\Lambda_{L1})}.
\label{eq:q}
\end{equation}
$Q$ gives a measure of the correlation between these two variables, since $Q=1$ for uncorrelated variables; $Q>1$ means positive correlation, while $Q<1$ means anti-correlation. The very high correlation between the two quantities in the top right corner suggests that, indeed, intense vortices have undergone strong stretching. For weak vortices, there are only very small positive (negative) correlations with small (large) amounts of stretching because, for those vortices,  viscous damping is strong relative to stretching.

\section{Summary}
\label{sec:summary}
We used the results of a direct numerical simulation of turbulence to study Lagrangian stretching by using the Cauchy-Green strain tensors. We have shown that the eigenvectors of the left and right Cauchy-Green tensors give a natural basis for studying phenomena involving stretching. In this paper, we have demonstrated this idea using the alignment statistics of two vectors: rod-like particles (essentially material-line segments) and the vorticity vector. Both rods and the vorticity vector tend to be aligned with the largest Lagrangian stretching direction $\hat{\bf{e}}_{L1}$, and the degree of alignment is stronger than it is with stretching directions defined from Eulerian quantities.

Rods become perfectly aligned with $\hat{\bf{e}}_{L1}$ in the long time limit. They rapidly become oriented in the plane $S_{12}$ formed by $\hat{\bf{e}}_{L1}$ and $\hat{\bf{e}}_{L2}$. However, it takes much longer for them to become perfectly aligned with $\hat{\bf{e}}_{L1}$ because a fraction of the rods experience nearly equal stretching in the $\hat{\bf{e}}_{L2}$ direction.


The stretching of vorticity, as an active vector, is usually studied in the Eulerian frame by using the alignment of vorticity with the eigenvectors of the instantaneous strain-rate tensor. 
Many studies have observed the puzzling result that vorticity tends to align most strongly with the intermediate eigenvector of the Eulerian strain rate. But after defining stretching in a Lagrangian basis, we find that the vorticity tends to align with the largest Lagrangian stretching direction, just as one would intuitively expect. In addition, the alignment of strong vorticity is almost exactly the same as for rods, and large vorticities are correlated with the strong stretching they have experienced. Analysis of Lagrangian stretching provides a powerful tool for understanding alignment of material lines and  vorticity in turbulence. These tools have the potential to illuminate many other problems including turbulent mixing, the dynamics of  anisotropic particles with other shapes than thin rods, and the structure of the events responsible for internal intermittency. 
\\

We thank Federico Toschi and Enrico Calzavarini for providing us with the DNS data.  We acknowledge support from US NSF grants DMR-1206399 to Yale University and DMR-1208990 to Wesleyan University, and COST Actions MP0806 and FP1005.

\bibliographystyle{jfm}

\end{document}